\documentclass[fleqn,usenatbib]{mnras}

\usepackage{newtxtext,newtxmath}

\usepackage[T1]{fontenc}

\DeclareRobustCommand{\VAN}[3]{#2}
\let\VANthebibliography\thebibliography
\def\thebibliography{\DeclareRobustCommand{\VAN}[3]{##3}\VANthebibliography}

\usepackage{graphicx}	
\usepackage{amsmath}	
\usepackage{longtable}
\defcitealias{2011ApJ...740...92G}{G11}
\defcitealias{2019ApJ...874...32R}{R19}
\defcitealias{2021MNRAS.503L..33Z}{Z21}

\title[Age Bias in SN Ia Cosmology. I.]{Strong progenitor age bias in supernova cosmology. I. Robust and ubiquitous evidence from a larger sample of host galaxies in a broader redshift range}

\author[C. Chung et al.]{
Chul Chung,$^{1}$\thanks{E-mail: chulchung@yonsei.ac.kr (CC)}
Seunghyun Park,$^{1}$
Junhyuk Son$^{1}$
Hyejeon Cho$^{1}$
and Young-Wook Lee$^{1}$\thanks{E-mail: ywlee2@yonsei.ac.kr (YWL)}
\\
$^{1}$Department of Astronomy \& Center for Galaxy Evolution Research, Yonsei University, Seoul 03722, Republic of Korea\\
}

\date{Accepted XXX. Received YYY; in original form ZZZ}

\pubyear{\the\year{}}

\begin{document}
\label{firstpage}
\pagerange{\pageref{firstpage}--\pageref{lastpage}}
\maketitle

\begin{abstract}
Type Ia supernovae (SNe Ia) serve as the most crucial standardizable candles in cosmology, providing direct measurements of the universe’s expansion history.
However, it is well-known that the post-standardization brightness of SNe Ia is influenced by the properties of their host galaxies, such as mass and star formation rate, both of which are closely related to progenitor age.
In this study, by measuring the stellar population ages of SN host galaxies, we reaffirm the ubiquitous and robust correlation between SN Ia luminosity and host age, showing that this host property dependence arises primarily from stellar population age of the host galaxy.
This analysis was conducted using an expanded sample of over 300 hosts across a broad redshift range up to $z\sim 0.4$, ensuring sufficient statistical significance of the result.
To quantify the relationship between host age and Hubble residual (HR), we employed two linear regression techniques: LINMIX, which assumes a Gaussian age error, and Bayesian hierarchical linear regression, which utilizes a full posterior for the age error.
Both models demonstrate a robust correlation between host age and HR, with high statistical significance approaching $5.5 \sigma$.
While our new regression analyses yield the slopes that are similar or slightly shallower compared to our previous results, the significance of these slopes has notably increased.
These findings robustly validate our previous suggestions that post-standardization SN Ia luminosity varies with progenitor age, which is currently not properly accounted for in SN cosmology.
\end{abstract}




\section{Introduction} \label{s1}

The empirically standardizable type Ia supernovae (SNe Ia) significantly enhance our understanding of cosmology by directly providing insights into the expansion history of the universe \citep{1998AJ....116.1009R, 1998ApJ...507...46S, 1999ApJ...517..565P}.
If SNe Ia can be effectively utilized as age-independent indicators of luminosity distance, they would provide reliable parameters that underpin cosmological models.
Numerous studies have leveraged these assumptions to predict cosmological parameters \citep[e.g.,][]{2022ApJ...938..110B, 2023arXiv231112098R, 2020A&A...641A...6P, 2024arXiv240403002D}.
However, it is widely recognized that systematic biases exist in the luminosity standardization of SNe Ia, previously believed to be determined solely based on their light curve width $x1$ and color parameter $c$ \citep[e.g.,][]{1993ApJ...413L.105P, 1998A&A...331..815T, 2005A&A...443..781G, 2007A&A...466...11G}.

The properties of host galaxy, especially stellar mass, are apparently correlated with the post-standardization brightness of SN Ia.
This host ``mass step'' \citep[e.g.,][]{2010ApJ...715..743K, 2010MNRAS.406..782S} is empirically corrected in the current practice of the SN Ia luminosity standardization process \citep[e.g.,][]{2014A&A...568A..22B, 2020MNRAS.494.4426S, 2021ApJ...923..237J}.
Introducing the host mass step is essential to the standardization process and has been fundamental since the initial discovery of the phenomenon.
Several recent studies \citep[e.g.,][]{2021ApJ...909...26B, 2023MNRAS.520.6214W} suggest that the brightness variations of SNe would be predominantly due to the variations in the ratio of total to selective extinction ($R_V$) rather than some direct effect from host galaxy's mass.
Nonetheless, the application of different $R_V$ values is still based on the host mass step, while their mass dependent $R_V$ values are in serious conflict with the observations \citep[e.g.,][]{2018ApJ...859...11S, 2024arXiv240606215P}.

Since the stellar population age of a host galaxy is closely linked with various parameters such as mass, metallicity, and dust content, the correlations between a supernova's luminosity (after standardization, unless otherwise specified throughout the paper) and host galaxy properties underscore the potential importance of age as the central parameter influencing these relationships. 
Initially overlooked as a critical factor, the age of the host galaxy has since emerged as a key determinant \citep{2016ApJS..223....7K, 2020ApJ...889....8K, 2020ApJ...903...22L, 2021MNRAS.503L..33Z, 2022MNRAS.517.2697L, 2023SCPMA..6629511W}.
Furthermore, \citet{2023ApJ...959...94C} recently demonstrated a bimodal age distribution of host galaxies through empirical correlations between galaxy mass and color.
Based on these, they reaffirm the root cause of the host mass step by demonstrating that the ages of host galaxies significantly influence the luminosities of SNe Ia.
Qualitatively, this aligns with \citet{2020A&A...644A.176R} and \citet{2024ApJ...961..185L}, who emphasized stellar population age as a key driver of SN Ia luminosity.
However, there is still a need to expand the sample size and extend the redshift range to fully validate the ubiquitous relationship between host age and SN Ia luminosity.

\begin{figure*}
\centering
\includegraphics[angle=-90,scale=0.7]{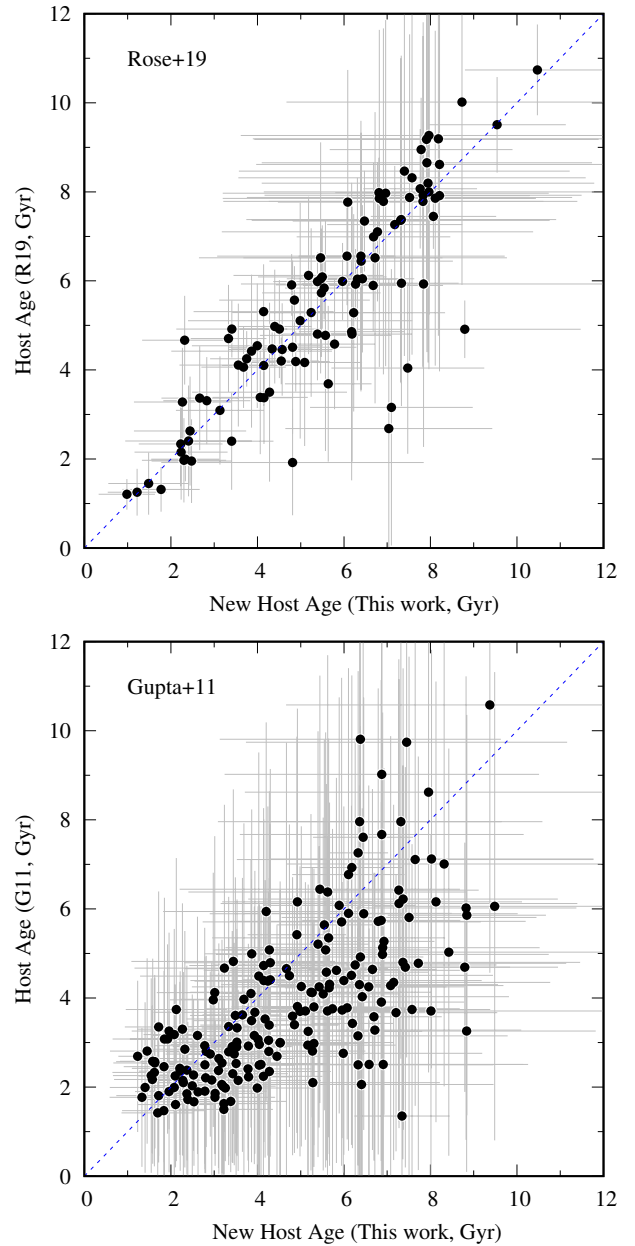}
\caption{
Comparison of host ages from \citetalias{2011ApJ...740...92G} and \citetalias{2019ApJ...874...32R} with our new age measurements.
The left and right panels are \citetalias{2011ApJ...740...92G} and \citetalias{2019ApJ...874...32R} samples, respectively.
Blue dashed lines represent the one-to-one age correlations.
Black circles are the median ages derived from the age posteriors of MCMC sampling, while the grey $1\sigma$ error bars represent the average range, based on the lower 13.6th and upper 86.4th percentiles of the age posterior distribution.
The upper panels of each plot show the age difference ($\Delta {\rm Host\, age}={\rm New\, Host\, Age} - {\rm Previous\, Host\, Age}$) as a function of ${\rm New\, Host\, Age}$.
The hosts from \citetalias{2019ApJ...874...32R} well align with the one-to-one correlation, albeit with some outliers.
In contrast, the host ages for \citetalias{2011ApJ...740...92G} exhibit larger deviations from this correlation and tend to show slightly older ages for the new age measurements.
}
\label{f1}
\end{figure*}

In this paper, we focus on the empirical relationship between the stellar population age of the host galaxy and the standardized SN Ia luminosity.
With the most updated stellar population models of \citet{2010ApJ...712..833C}, we measure population ages of SN Ia host galaxies from the samples of \citet{2011ApJ...740...92G} and \citet{2019ApJ...874...32R}, using a consistent age-dating framework described in Sections~\ref{s2}-\ref{s4}.
Following these new age measurements, we explore the statistical significance of the relationship between the age of the host galaxy and the luminosity of SN Ia in Sections~\ref{s5}-\ref{s7}.
Finally, discussions related to the impact of the age bias are presented in Section~\ref{s8}.

\section{Host Galaxy Sample} \label{s2}

We have substantially increased the sample of host galaxies with robustly measured stellar population ages compared to previous studies.
Our primary host galaxy sample is drawn from the host galaxies of \citet[][hereafter \citetalias{2011ApJ...740...92G}]{2011ApJ...740...92G}, which consist of 206 host galaxies with a redshift range up to $\sim 0.4$.
The extensive sample of \citetalias{2011ApJ...740...92G} host galaxies allow for a more comprehensive analysis and provide a greater statistical significance.
Since the host age measurements by \citetalias{2011ApJ...740...92G}, significant updates have been made to the population synthesis models \citep{2009ApJ...699..486C, 2010ApJ...712..833C} used for age determination, along with several important technical advances such as more precise age estimation using Markov Chain Monte Carlo (MCMC) methods and improved star formation history (SFH) prescriptions.
These developments require new age measurements for the \citetalias{2011ApJ...740...92G} sample.
To compare our age measurements with other studies, we further employed the \citet[][hereafter \citetalias{2019ApJ...874...32R}]{2019ApJ...874...32R} host galaxies, which are also drawn from SDSS.
For the hosts in \citetalias{2011ApJ...740...92G}, we obtained the \(ugriz\) magnitudes from the SDSS DR16\footnote{https://skyserver.sdss.org/dr16/} database \citep{SDSS_DR16} using the coordinates provided in \citetalias{2011ApJ...740...92G}, while for the \citetalias{2019ApJ...874...32R} host galaxies, we directly adopted the $ugriz$ magnitudes from Table~3 in \citetalias{2019ApJ...874...32R}.
There are 33 overlapping hosts between \citetalias{2011ApJ...740...92G} and \citetalias{2019ApJ...874...32R}.
Upon examining the magnitude differences in these hosts between \citetalias{2011ApJ...740...92G} and \citetalias{2019ApJ...874...32R}, we found the discrepancies to be negligible.
Among the 220 hosts from \citetalias{2011ApJ...740...92G}, we measured ages for 199 hosts with available SDSS DR16 magnitudes, and for \citetalias{2019ApJ...874...32R}, we used all of the 102 hosts for the new age measurements.

In many studies of SN Ia host galaxies, the Hubble residual (HR) at a given redshift serves as a measure of relative SN Ia luminosity and is commonly used to assess how host properties affect the brightness of SN Ia.
This is based on the assumption that any deviation from the cosmic expansion model in a Hubble diagram is due to an incorrectly assumed luminosity of SNe, and the residuals can be used to find relationships with host properties. 
Other potential effects could also contribute to HRs, such as a progenitor property deviating from the mean/average host property and the peculiar velocity of the host galaxy \citep[see, e.g.,][]{2014A&A...568A..22B}, and some of the scatter observed in the HR diagrams may be attributed to these effects.
When employing HRs, we used the same HR values as those adopted in the \citetalias{2011ApJ...740...92G} and \citetalias{2019ApJ...874...32R}, respectively, rather than incorporating other catalogs.
The light-curve analysis for the HRs in \citetalias{2011ApJ...740...92G} was conducted using the SNANA package \citep{2009PASP..121.1028K}, following the criteria adopted in \citet{2009ApJS..185...32K}.
For the hosts in \citetalias{2019ApJ...874...32R}, HRs were obtained from the HR values reported in \citet{2013ApJ...763...88C}, which passed the cosmology selection criteria.
Importantly, neither analysis applied mass step corrections to the HRs.
This is essential in our analysis as the mass step correction can partially mitigate the age dependence of the HR when the sample is in a similar redshift bin.
Note that in our analysis of the correlation between HR and host galaxy age, we examined the two samples separately to maintain the consistency of their distinct characteristics.
For the \citetalias{2011ApJ...740...92G} sample, which spans a large redshift range, we applied additional corrections to the original HRs to account for the redshift dependence resulting from the assumed baseline cosmology and cosmological age evolution.
A detailed description of this process is provided in Section~\ref{s4}.

\begin{figure}
\centering
\includegraphics[angle=0,scale=0.7]{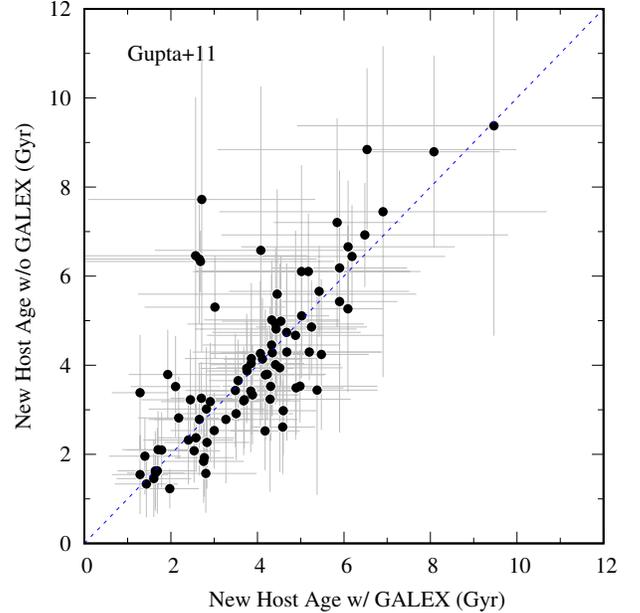}
\caption{
Comparison of host ages measured with and without the inclusion of GALEX FUV and NUV photometry for \citetalias{2011ApJ...740...92G} sample.
The upper panel shows the age measurement deviation without GALEX compared to the age measured with GALEX.
While most hosts align with the one-to-one correlation, a few are estimated to be younger when including GALEX data.
}
\label{f2}
\end{figure}

\section{New Age measurement}\label{s3}

Age dating of stellar populations based on absorption features in spectra is the most reliable method among various age-dating techniques \citep{1994ApJS...95..107W, 2011MNRAS.412.2183T, 2013ApJS..204....3C}.
However, this approach encounters challenges and limitations when applied to galaxies with recent star formation (RSF).
In most late-type galaxies, where emission lines from RSF dominate over absorption lines, correcting for emission to accurately measure absorption line strengths becomes highly uncertain.
Furthermore, reliable spectroscopic observations of absorption lines are difficult at larger distances.
Although photometric magnitudes for age dating face similar limitations, photometric age determination, which primarily relies on overall continuum differences rather than specific absorption features, mitigates these challenges to some extent.
Thus, in this study, we are compelled to use photometric age dating to determine consistent ages of host galaxies across diverse morphological types and broad redshift ranges.

An additional point to consider when performing age dating based on spectro-photometric observations for galaxies with RSF is that luminosity-weighted age (LWA) derived from simple stellar population models is significantly biased by a small number of hot, young stars, which strongly influence the shape of spectral energy distributions \citep{2013ApJS..204....3C, 2020ApJ...889....8K}.
To obtain reliable age estimates applicable to all morphological types of galaxies with RSF, mass-weighted age (MWA), which fully accounts for their SFH, should be used instead \citep{2020ApJ...903...22L}.
In this context, despite the larger uncertainties associated with photometric observations, the population model of \citet{2010ApJ...712..833C} remains the most effective tool for estimating SFH and MWA based on photometric data.
Some SN Ia progenitors may have ages that deviate from the MWAs of their host galaxies. 
However, as shown in \citet{2014MNRAS.445.1898C} and \citet[][see their Figure~6]{2022MNRAS.517.2697L}, models incorporating the delay-time distribution of SNe Ia and the cosmic star formation history suggest that, on average, the age differences between the stellar populations and SN Ia progenitors in host galaxies are not significant. 
Notably, the relative age difference ($\Delta$age) as a function of redshift, which is crucial for cosmological implications, also follows a similar trend.

The previous age measurements by \citetalias{2011ApJ...740...92G} relied on a very early version of the Flexible Stellar Population Synthesis (FSPS) code by \citet{2009ApJ...699..486C} and \citet{2010ApJ...712..833C}, which incorporated the Padova isochrones \citep{2007A&A...469..239M, 2008A&A...482..883M} and the BaSeL 3.1 spectral library \citep{1997A&AS..125..229L, 2002A&A...381..524W}. 
\citetalias{2011ApJ...740...92G} generated four grids of FSPS parameters, including the $e$-folding time scale of SFH and the start time of star formation, and derived the MWA of each host galaxy from these grids.
Here, we adopt the updated FSPS version 0.4.6\footnote{https://pypi.org/project/fsps/0.4.6/} (as of October 2023), with new age-dating approach that differs significantly from \citetalias{2011ApJ...740...92G}. 
Notable updates to the FSPS model include the use of the MIST isochrones  \citep{2016ApJ...823..102C, 2016ApJS..222....8D} and the MILES spectral energy distributions \citep{2011A&A...532A..95F}, which introduce a revised age scale compared to that in \citetalias{2011ApJ...740...92G}. 
Additionally, the application of MCMC methods with updated SFH prescriptions, which incorporates both early burst and recent star formation, refines the age measurements of \citetalias{2011ApJ...740...92G} host galaxies.
In practice, we follow the methodology of \citetalias{2019ApJ...874...32R}\footnote{https://github.com/benjaminrose/mc-age} to remeasure the ages of approximately 300 host galaxies from both \citetalias{2011ApJ...740...92G} and \citetalias{2019ApJ...874...32R}.
Within the FSPS model, spectral energy distributions are shifted to the corresponding redshift to account for the observed host magnitudes of \citetalias{2011ApJ...740...92G} and \citetalias{2019ApJ...874...32R}, which span a broad redshift range.
To maintain consistency in the age scale of \citetalias{2011ApJ...740...92G} and \citetalias{2019ApJ...874...32R} host galaxies, we applied the same FSPS model settings for metallicity, dust, and SFH used in \citetalias{2019ApJ...874...32R}.
The new ages were estimated using MCMC sampling with the same priors applied in \citetalias{2019ApJ...874...32R}.
For a detailed description of the MCMC age sampling methodology, see Sections~3-4 of \citetalias{2019ApJ...874...32R}.

\begin{table}
\begin{center}
\caption{\label{t1} New age measurements for \citetalias{2011ApJ...740...92G} hosts}
\begin{tabular}{ccccc} 
\hline
{SNID} & {Host Age} & {$\sigma_{\rm Age}$} & {Hubble Residual} & {$\sigma_{\rm HR}$}\\
 & (Gyr)& (Gyr)& (mag)& (mag)\\
\hline
 1166 & 5.24 & 2.79 & -0.27 (-0.33) & 0.22 \\
 1253 & 6.46 & 3.56 & -0.10 (-0.11) & 0.16 \\
 1371 & 6.92 & 1.17 & -0.24 (-0.18) & 0.06 \\
 1580 & 6.70 & 1.57 & -0.15 (-0.12) & 0.08 \\
 1688 & 3.23 & 1.13 & 0.05 (0.02) & 0.23 \\
\hline
\multicolumn{5}{p{8cm}}{Note. The complete Table~\ref{t1} is available as online material. The SNID in Tables~\ref{t1} and \ref{t2} corresponds to the IDs from the SDSS-II Supernova Survey \citep{2008AJ....135..348S}.} 
\end{tabular}
\end{center}
\end{table}

\begin{table}
\centering
\caption{New age measurements for \citetalias{2019ApJ...874...32R} hosts}
\label{t2}
\begin{tabular}{ccccc} 
\hline
{SNID} & {Host Age} & {$\sigma_{\rm Age}$} & {Hubble Residual} & {$\sigma_{\rm HR}$}\\
 & (Gyr)& (Gyr)& (mag)& (mag)\\
\hline
 762 & 6.96 & 3.77 &   0.15 & 0.08 \\
 1032 & 8.21 & 4.23 &  -0.15 & 0.12 \\
 1371 & 7.76 & 1.31 &  -0.14 & 0.06 \\
 1794 & 2.48 & 0.92 &   0.27 & 0.08 \\
 2372 & 6.19 & 1.39 &  -0.12 & 0.07 \\
\hline
\multicolumn{5}{p{8cm}}{Note. The complete Table~\ref{t2} is available as online material.} 
\end{tabular}
\end{table}

Tables~\ref{t1} and \ref{t2} provide a portion of our new age measurements for the sample galaxies associated with each SN Ia.
Figure~\ref{f1} shows a one-to-one comparison of the age scales with the previously reported ages in \citetalias{2011ApJ...740...92G} and \citetalias{2019ApJ...874...32R} along with the age difference ($\Delta$ Host age) relative to our new host age.
While the newly measured ages for the \citetalias{2019ApJ...874...32R} sample generally agree with the previously measured ages in \citetalias{2019ApJ...874...32R}, those for the \citetalias{2011ApJ...740...92G} sample exhibit significant shifts compared to the previous measurements in \citetalias{2011ApJ...740...92G}.
Given that the correlation between host age and HR is a critical factor in identifying the age bias in SN cosmology \citep[e.g.,][]{2020ApJ...889....8K, 2020ApJ...903...22L, 2022MNRAS.517.2697L}, uncovering the differences in age from previous results is a crucial aspect of this study.

Considering that only an updated version of FSPS was applied while the age-dating method remained unchanged in our analysis, the strong correlation observed for \citetalias{2019ApJ...874...32R} hosts is a natural and expected outcome.
In contrast, the systematic shift observed in the \citetalias{2011ApJ...740...92G} sample is likely due to the advances both in the FSPS and in the age-dating methods.
While \citetalias{2011ApJ...740...92G} employed $\chi^2$ minimization, which estimates host ages based on fixed model grids (see Table 1 of \citetalias{2011ApJ...740...92G}), our method utilizes an MCMC sampling, enabling us to determine the most probable host age without being constrained by the model grids.
Beyond overcoming the grid resolution limitations of $\chi^2$ minimization, the MCMC method is more robust, as it marginalizes nuisance parameters and explores the full posterior distribution, yielding more accurate uncertainty estimates while reducing biases from local minima and Gaussian assumptions.
This, combined with the updated version of FSPS, is expected to result in significant differences in age estimation.
Moreover, as \citetalias{2019ApJ...874...32R} reported inconsistencies arising from the use of new SFH models, the age shift observed in our analysis, where younger hosts appear slightly younger and older hosts appear slightly older compared to \citetalias{2011ApJ...740...92G} ages (see upper left panel of Figure~\ref{f1}), can similarly be attributed to these improved SFH models.
These combined effects can explain the observed shifts between the newly measured and previously reported \citetalias{2011ApJ...740...92G} host ages.

The other factor potentially contributing to the larger median age deviation in \citetalias{2011ApJ...740...92G} hosts is the different use of passbands between our work and the previous \citetalias{2011ApJ...740...92G} analysis.
In \citetalias{2011ApJ...740...92G} analysis, for some host galaxy samples, they added GALEX FUV and NUV photometry to the SDSS photometric data.
In general, bright FUV and NUV magnitudes indicate either a strong star formation rate (SFR) \citep[e.g.,][]{2005ApJ...619L...1M, 2023ApJS..268...26O} or extremely hot and old stellar components in the object \citep[e.g.,][]{1997ApJ...476...28P, 2011ApJ...740L..45C, 2024ApJ...966...50A}.
Therefore, although incorporating GALEX photometry with FUV and NUV filters might improve an age estimation accuracy, \citetalias{2011ApJ...740...92G} noted that adding GALEX data to SDSS primarily enhances the determination of dust optical depth and SFR \citep{2005ApJ...619L..39S}.
In Figure~\ref{f2}, we compare the measured ages of 87 \citetalias{2011ApJ...740...92G} hosts with available GALEX FUV and NUV magnitudes,\footnote{https://galex.stsci.edu/casjobs/} both including and excluding the GALEX photometry in the MCMC age sampling.
If there are noticeable discrepancies between the two age scales, it might indicate the impact of including GALEX photometry in the age estimation.
However, most of the hosts align well with the age scale derived without GALEX data.
Some outliers appear to have younger ages when the GALEX data are included, and these galaxies may be related to the UV-upturn phenomenon \citep{2011ApJ...740L..45C}, although confirmation is challenging with the limited passbands available.
Overall, we conclude that, due to the minimal impact of GALEX photometry, host ages based on SDSS photometric data are sufficiently reliable for further analysis.

\section{Empirical Correction of Hubble Residual for Redshift Evolution} \label{s4}

\begin{figure*}
\centering
\includegraphics[angle=0,scale=0.7]{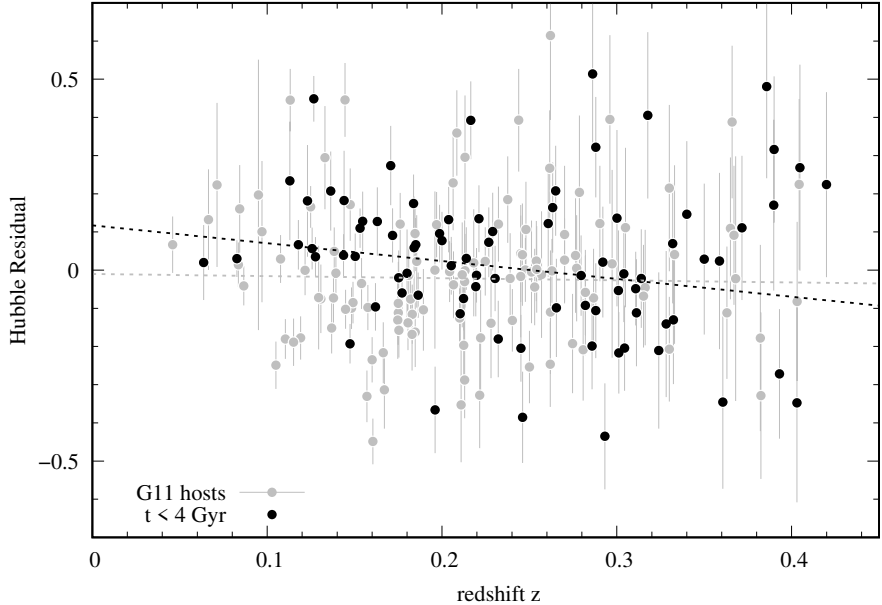}
\caption{
Hubble residual (HR) distribution as a function of redshift in \citetalias{2011ApJ...740...92G} sample.
Host galaxies younger than 4 Gyr are marked as filled black circles.
Linear regressions for all hosts and younger hosts are shown as grey and black dashed lines, respectively.
The homogeneous younger age group exhibits systematically different HRs with respect to redshift.
}
\label{f3}
\end{figure*}

In SN cosmology, HR at a given redshift is used to estimate the relative luminosity of SNe Ia.
The value of HR is defined as the difference in distance modulus between the SN Ia and the chosen baseline cosmological model (${\rm HR}(z) = \mu_{\rm SN} - \mu_{\rm model}(z)$).
Under a well-fitted baseline cosmology, the average HR ideally remains around ${\rm HR}\sim 0.0$. 
However, if the baseline cosmological model is incorrectly determined, SNe Ia of the same progenitor age at different redshifts can display varying HR values due to inaccurate distance modulus predictions.
Additionally, cosmological age evolution changes the progenitor age distributions of SNe Ia across redshifts, which further affects redshift-dependent HR values because of population age differences.
To mitigate both effects, we apply a correction to the HR value so that SNe Ia of the same age maintain consistent HR values regardless of redshift. 
For the \citetalias{2019ApJ...874...32R} sample, where host galaxies are confined to a narrow redshift range around $ z \sim 0.14 $, the HR variation attributable to redshift is negligible.
In contrast, the \citetalias{2011ApJ...740...92G} host sample, which spans up to $ z \sim 0.4 $, exhibits redshift differences that can lead to significant HR variations.

To deal with cosmological age evolution of HR values with respect to redshift, we select 82 relatively young hosts ($<4$~Gyr) from the \citetalias{2011ApJ...740...92G} sample, as their HRs are less influenced by cosmological age differences across redshifts.
In Figure~\ref{f3}, we examine the HR evolution of \citetalias{2011ApJ...740...92G} hosts with respect to redshift.
The linear regressions for all hosts and younger hosts ($< 4$~Gyr) are ${\rm HR}_{\rm all} = (-0.055 \pm 0.181)z - (0.010 \pm 0.035)$ and ${\rm HR}_{\rm young} = (-0.467 \pm 0.256)z + (0.117 \pm 0.052)$, respectively.
The HR values of the overall sample is distributed around ${\rm HR}\sim 0.0$, with a negligible slope relative to redshift.
However, the distribution of HRs for the young hosts across the entire redshift range exhibits a slope deviating from the overall sample.
Since the SNe Ia of the same age must maintain consistent HR values regardless of redshift, this indicates that a correction is necessary as a function of redshift.
Therefore, we applied adjustments to the HRs of all \citetalias{2011ApJ...740...92G} hosts using the offset of the regression line for the young age group from the ${\rm HR}=0.0$.
The corrected HR values for \citetalias{2011ApJ...740...92G} hosts are provided in Table~\ref{t1}, with the original \citetalias{2011ApJ...740...92G} values in the brackets.

\begin{figure*}
\centering
\includegraphics[angle=0,scale=0.8]{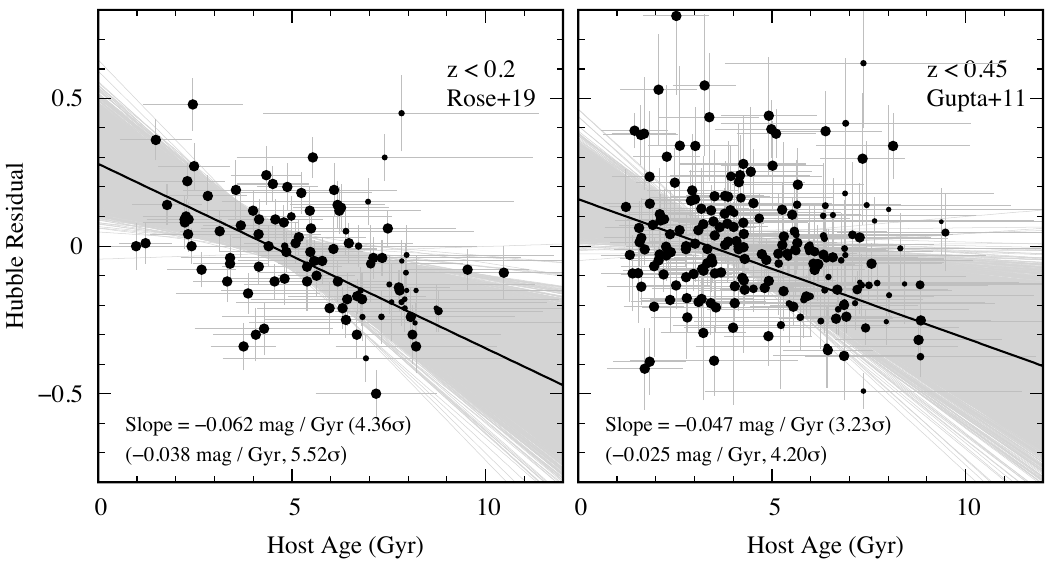}
\caption{
The LINMIX results for \citetalias{2019ApJ...874...32R} and \citetalias{2011ApJ...740...92G} host galaxies.
The gray lines are the LINMIX simulation for 10,000 regressions.
The thick black lines indicate the slopes based on the likelihood distribution of the LINMIX analysis.
The size of the points inversely reflects the age measurement errors.
For comparison, the slope and $\sigma$ value, derived from full posterior for the host age distribution using Bayesian hierarchical linear regression (see Section~\ref{s6}), are also provided in parentheses.
}
\label{f4}
\end{figure*}

\begin{figure*}
\centering
\includegraphics[angle=-90,scale=0.69]{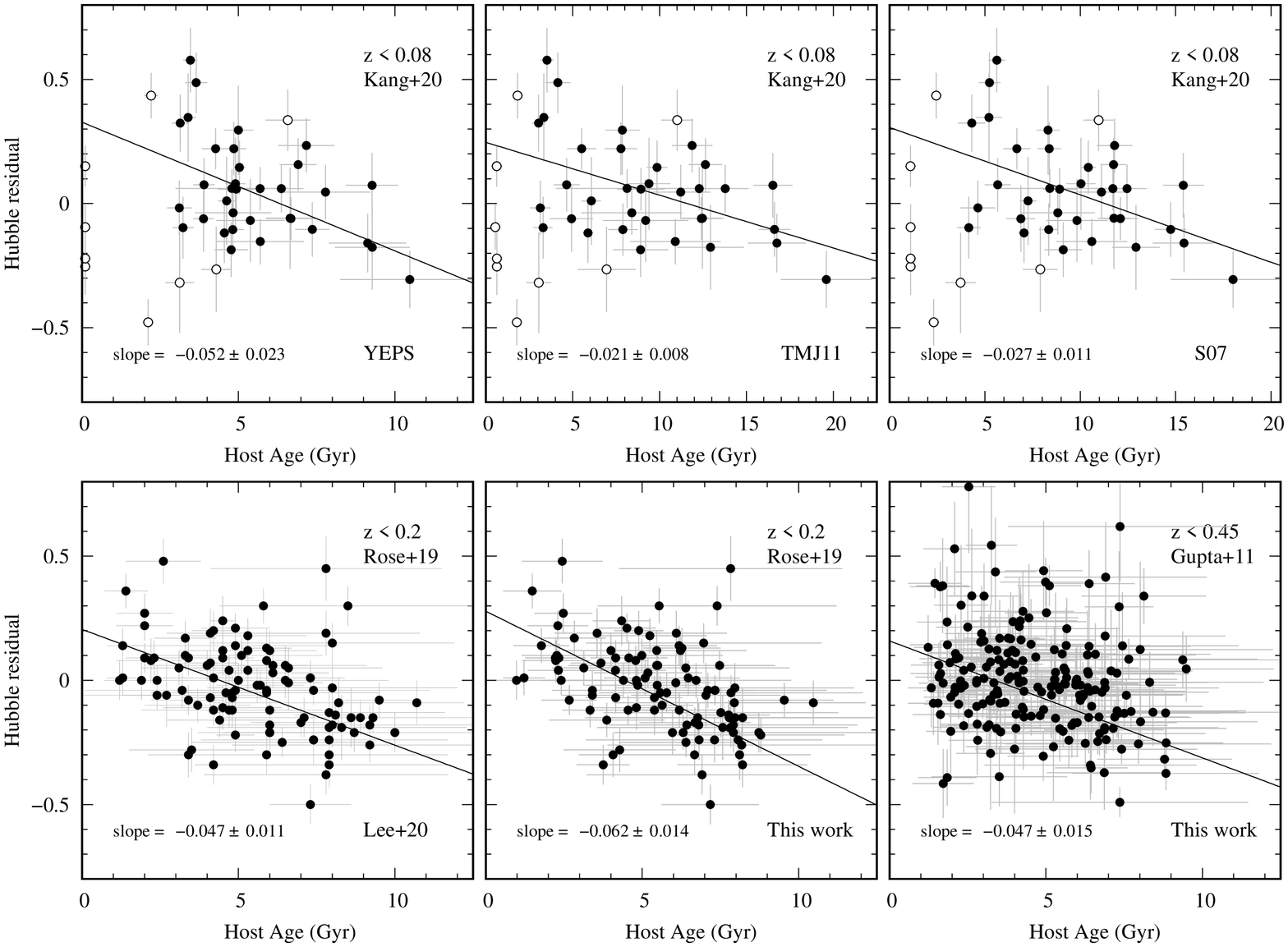}
\caption{
Correlation between host age and HR for \citet{2020ApJ...889....8K}, \citetalias{2019ApJ...874...32R}, and \citetalias{2011ApJ...740...92G} host samples.
The redshift range for each sample is denoted at the upper right corner of each panel.
As in Figure~\ref{f4}, the point size is inversely proportional to the age measurement errors.
Top row: Early-type host ages from \citet{2020ApJ...889....8K} estimated by three different population synthesis models, YEPS \citep{2013ApJS..204....3C}, TMJ11 \citep{2011MNRAS.412.2183T}, and \citet{2007ApJS..171..146S}, respectively.
Hosts with star formation signatures (indicated as open circles) are not included in the regression analysis (see the text).
Bottom row: The left panel is our earlier analysis based on the \citetalias{2019ApJ...874...32R} data \citep{2020ApJ...903...22L}, while the middle and right panels show our new analysis in Figure~\ref{f4} for \citetalias{2019ApJ...874...32R} and \citetalias{2011ApJ...740...92G} samples.
All three host samples show consistent correlations between host age and SN Ia luminosity based on the LINMIX regression.}
\label{f5}
\end{figure*}

\section{Progenitor age bias predicted from LINMIX analysis} \label{s5}

The most commonly used linear regression method in SN Ia age bias analysis \citep[\citetalias{2011ApJ...740...92G}; ][]{2014MNRAS.438.1391P, 2020MNRAS.491.5897P, 2020ApJ...889....8K, 2020ApJ...903...22L, 2022MNRAS.517.2697L, 2023SCPMA..6629511W} is the LINMIX package \citep{2007ApJ...665.1489K}.
The LINMIX implements an MCMC posterior sampling method, enabling linear regression predictions for data with measurement errors in both variables.
It assumes that measurement errors in both variables follow a normal Gaussian distribution and generates regression models by applying multiple fits to resampled data subset using Bayesian inference.
Since a full posterior for the age distribution of a specific host does not necessarily follow a typical normal Gaussian distribution, this approach may not always be the most suitable for deriving the general correlation between host age and HR \citep[see][]{2021MNRAS.503L..33Z}.
Nevertheless, while it may have some limitations, using the median host age and its 13.6th to 86.4th percentile range from the full posterior distribution as the average $1\sigma$ error provides a generally robust and effective summary of the representative values for host stellar populations.

Figure~\ref{f4} shows the LINMIX results for the newly measured median host ages versus HRs.
This plot includes the LINMIX simulation for 10,000 regressions, represented by gray lines.
This analysis suggests slopes of $-0.062 \pm 0.014$ ($4.36\sigma$) and $-0.047 \pm 0.015$\footnote{The age bias for the \citetalias{2011ApJ...740...92G} sample, derived using the LINMIX analysis without the HR correction in Section~\ref{s4}, still yields $-0.037 \pm 0.015$ based on our new age dating. Although the slope and significance are smaller compared to the case with the HR correction, the similar negative slope is once again confirmed.} ($3.23\sigma$) for the hosts in \citetalias{2019ApJ...874...32R} and \citetalias{2011ApJ...740...92G}, respectively.
Compared to our previously measured slope of $-0.047 \pm 0.011$ for \citetalias{2019ApJ...874...32R} hosts in \citet{2020ApJ...903...22L}, the regression based on new age measurements prefers a steeper slope.
As highlighted in Section~\ref{s3}, this is due to the application of the up-to-date version of FSPS and the changes in the measured ages.
For the \citetalias{2011ApJ...740...92G} hosts, the slope has increased from $-0.015 \pm 0.008$ to $-0.047 \pm 0.015$.
This significant change can be attributed to the updated age-dating method and the additional HR correction relative to redshift employed in this analysis (see Sections 3-4), with the MCMC age sampling improving the precision of age measurement for \citetalias{2011ApJ...740...92G} hosts.
With relatively well-determined host ages, a strong correlation between host age and HR has now become evident, which was not clear in the earlier study.

Despite this significant change from the original analysis of \citetalias{2011ApJ...740...92G}, our new result for \citetalias{2011ApJ...740...92G} sample shows a shallower slope compared to that of \citetalias{2019ApJ...874...32R} hosts.
This difference likely stems from mixing hosts across different redshifts, which have varying observational quality, especially at greater redshifts.
If we limit the \citetalias{2011ApJ...740...92G} hosts to those with $z<0.2$, where \citetalias{2019ApJ...874...32R} hosts are located, the slope becomes somewhat steeper ($-0.050 \pm 0.016$).
The probabilities of a negative slope, based on the LINMIX results for \citetalias{2019ApJ...874...32R} and \citetalias{2011ApJ...740...92G} hosts, are greater than 99.998\% and 99.937\%, respectively, indicating a significance level comparable to that of the slope itself.

In Figure~\ref{f5}, we compile and display the results of the age-HR correlation reported to date using the LINMIX method for all galaxy morphological types.
The top panels of the figure summarize our previous works for nearby early-type host galaxies ($z<0.08$).
The bottom panels are our new results for the R19 and G11 samples in Figure~\ref{f4}, together with our previous analysis for the R19 sample (Figure~1 from \citealt[][]{2020ApJ...903...22L}) for comparison.
The LINMIX regressions for the early-type hosts as reported in \citet{2020ApJ...889....8K} indicate slopes of $-0.052 \pm 0.023$ ($2.26\sigma$), $-0.021 \pm 0.008$ ($2.63\sigma$), and $-0.027 \pm 0.011$ ($2.45\sigma$), depending on the population synthesis models employed.
The reported slope for \citetalias{2019ApJ...874...32R} sample in \citet{2020ApJ...903...22L} is $-0.047 \pm 0.011$ ($4.27\sigma$), indicating that the age bias identified in previous studies consistently suggests a correlation between host age and HR.
Our new age measurements, which employed homogeneous passbands of $ugriz$, the up-to-date FSPS model, and a robust MCMC sampling method, further support our previous analyses, revealing a substantial age bias in 300 host galaxies with a significance level approaching $\sim 4.4\sigma$.

\begin{figure}
\centering
\includegraphics[angle=-90,scale=0.47]{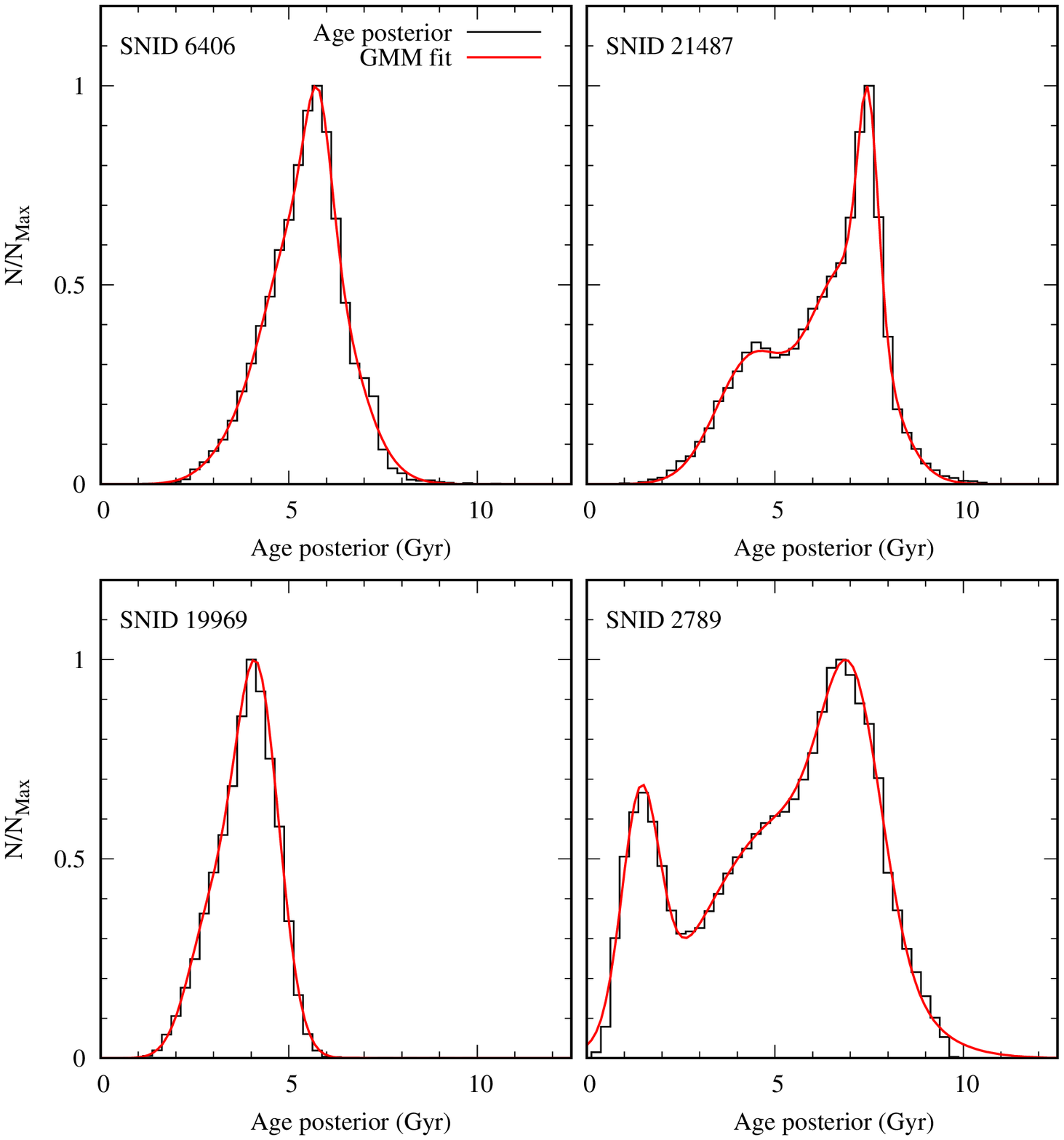}
\caption{
Examples of three-Gaussian mixture fitting to the full posterior distribution for host  age from MCMC sampling.
The upper and lower panels show host galaxies selected from \citetalias{2019ApJ...874...32R} and \citetalias{2011ApJ...740...92G} samples, respectively.
The typical cases with unimodal and bimodal distributions are shown in black histograms.
The red lines are three-Gaussian fitting results for each host.
Overall, the fits show good agreement with the original full posteriors.
}
\label{f6}
\end{figure}

\begin{figure}
\centering
\includegraphics[angle=0,scale=0.46]{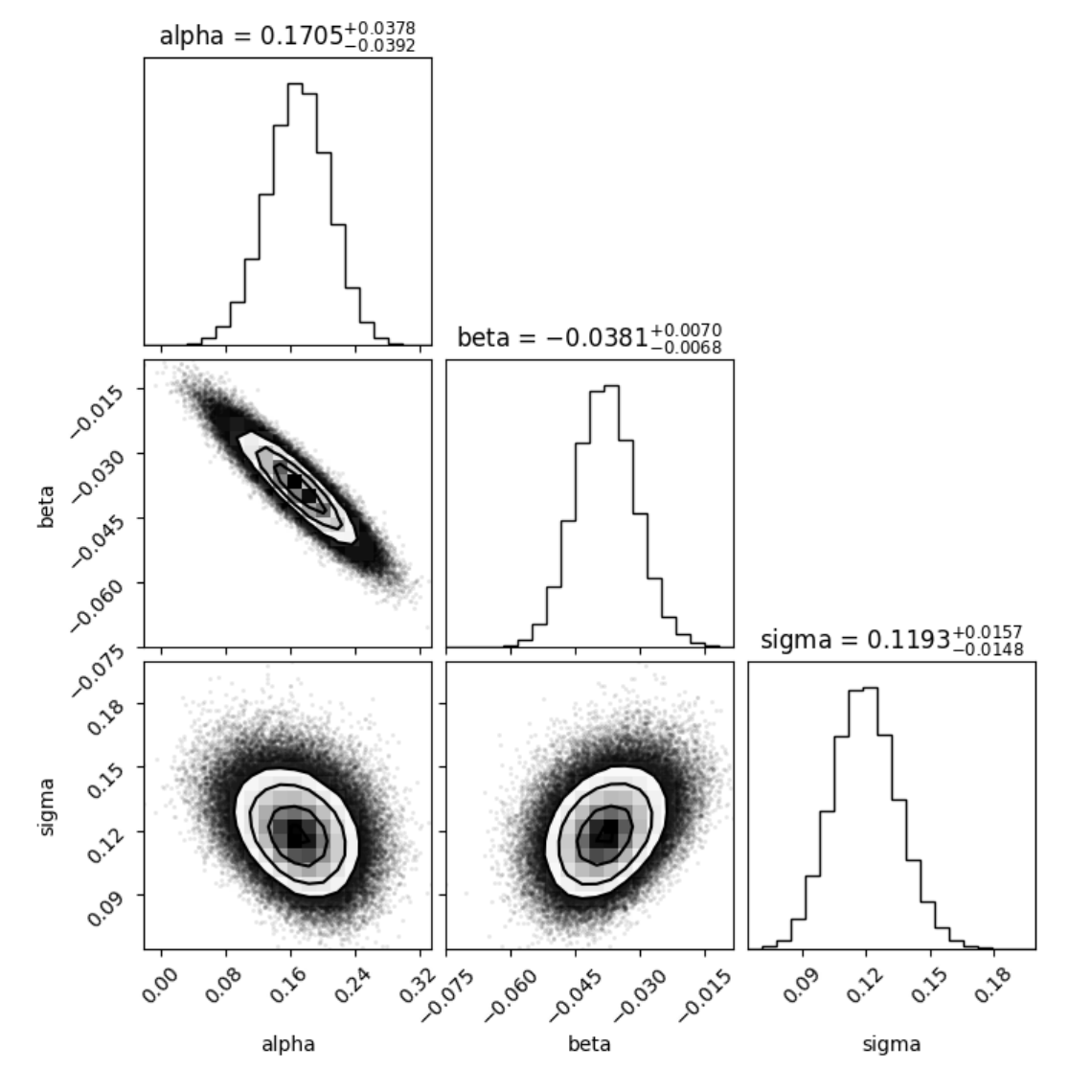}
\caption{
Posterior distributions for Bayesian hierarchical linear regression parameters for the \citetalias{2019ApJ...874...32R} sample.
The slope $\beta$ shows a significant correlation of $-0.038 \pm 0.007$ with $\sim 5.5\sigma$ significance.
}
\label{f7}
\end{figure}

\begin{figure}
\centering
\includegraphics[angle=0,scale=0.46]{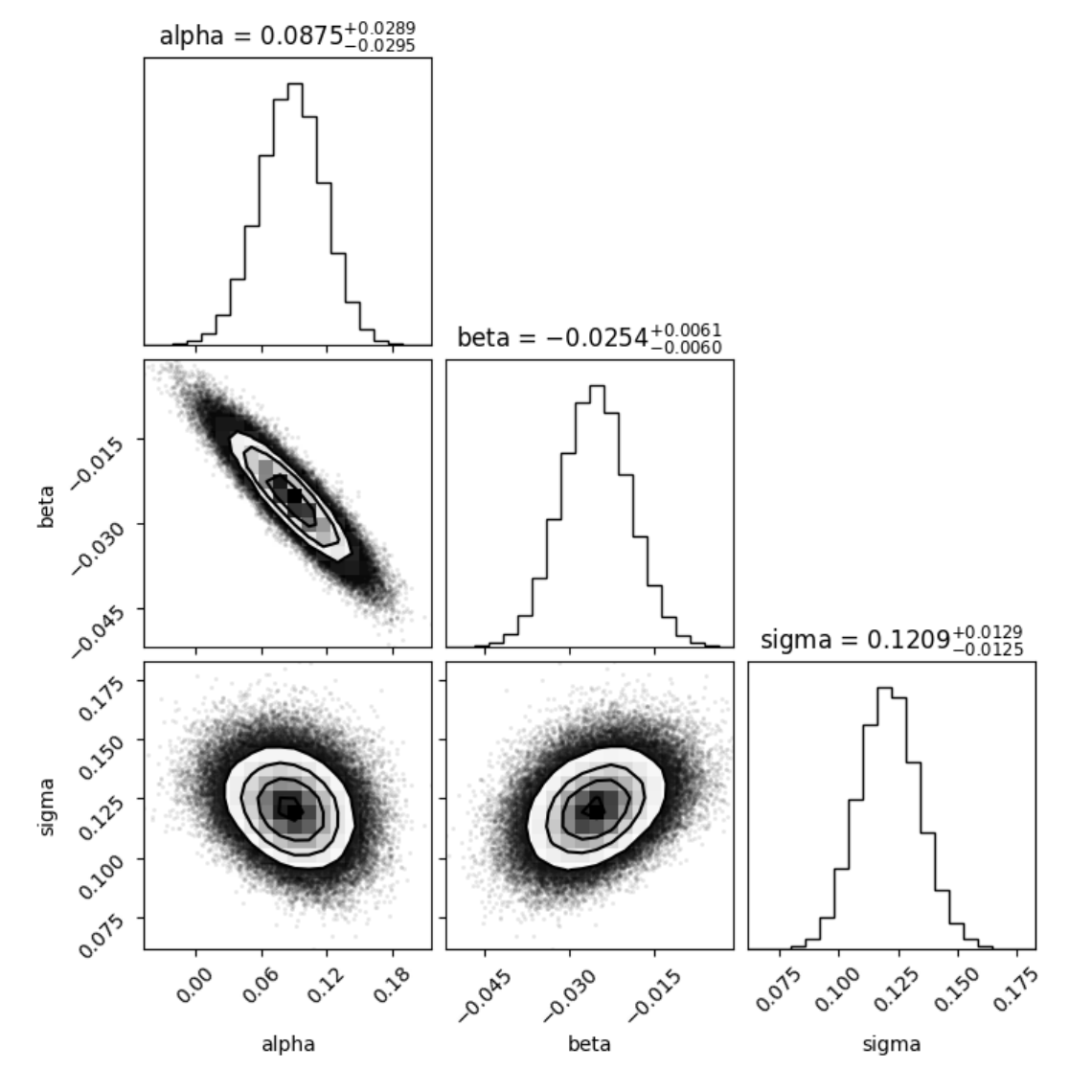}
\caption{
Same as Figure~\ref{f7} but for \citetalias{2011ApJ...740...92G} sample.
The slope $\beta$ and its significance is $-0.025 \pm 0.006$ at $\sim 4.2\sigma$.
}
\label{f8}
\end{figure}

\begin{figure*}
\centering
\includegraphics[angle=-90,scale=0.66]{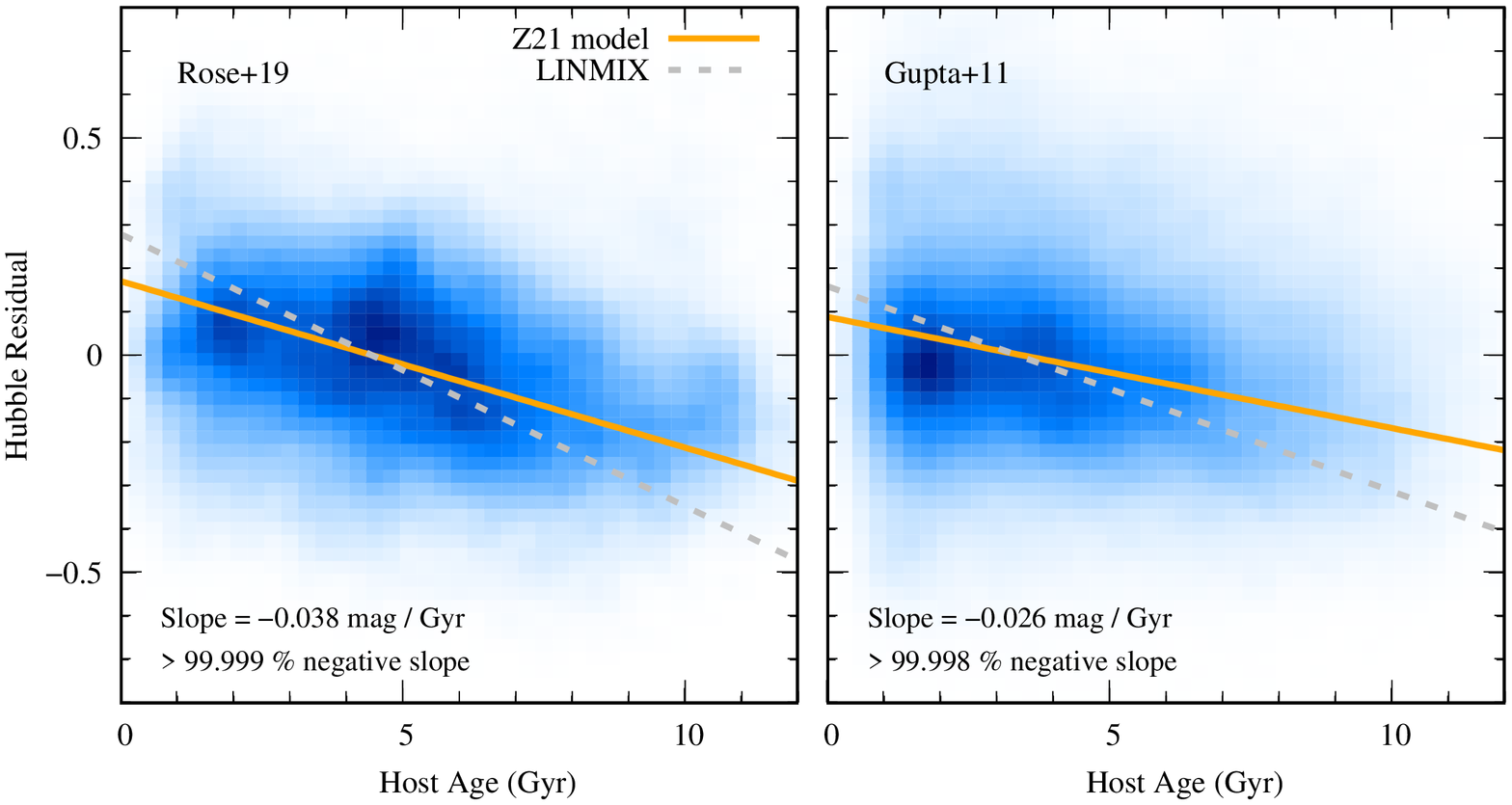}
\caption{
Comparison of the regression slopes predicted by \citetalias{2021MNRAS.503L..33Z} regression model with those from the LINMIX analysis.
The bluish density map is derived from 1,000 Monte Carlo random samples of each SN Ia host based on its HR and full posterior distribution for host age.
The solid orange lines and dashed gray lines are the results from \citetalias{2021MNRAS.503L..33Z} regression model and the LINMIX, respectively.
The probability of a negative slope for each host sample is indicated at the bottom of each panel.
}
\label{f9}
\end{figure*}

\section{Progenitor age bias predicted from the full posterior distributions for host ages}\label{s6}

In Section~\ref{s5}, we employed the LINMIX package to derive the linear regression between the population ages of host galaxies and the HRs of SNe in them, based on the median value of the full posterior distribution for host age, with the 13.6th and 86.4th percentile range representing the average scatter for the normal Gaussian age distribution.
However, this approach may not fully reflect the true age distributions, as not all posterior distributions for host ages adhere to a single normal distribution; instead, they occasionally exhibit broader, bimodal distributions.
Therefore, we test the linear regression model using a Bayesian hierarchical approach, incorporating the full posterior distribution for host age rather than assuming the normal Gaussian host age distribution employed in the LINMIX method.
Although this approach can cause a dilution bias, leading to the attenuation of the regression \citep[e.g.,][]{2020ApJ...903...22L}, we employ the method proposed by \citet[][hereafter \citetalias{2021MNRAS.503L..33Z}]{2021MNRAS.503L..33Z} to minimize the potential issues arising from a misrepresented posterior age distribution.

Following \citetalias{2021MNRAS.503L..33Z}, we construct a Bayesian hierarchical linear regression model of the form: $ y = \alpha + \beta x + \epsilon_{\rm int} $, where $\alpha$ represents the intercept, $\beta$ the slope, and $\epsilon_{\rm int} \sim N(0, \sigma_{\rm int}^2)$ the scatter, modeled as a normal distribution with intrinsic scatter $\sigma_{\rm int}$.
We first fit the full posterior distribution for host ages using a simple three-Gaussian mixture model.
Figure~\ref{f6} presents examples of this fitting.
There are some small differences in the case of broad posterior distributions for host ages due to the limited number of Gaussians we adopted in the analysis, but most cases show good fits with three-Gaussians.
The fitting parameters for a given host $i$ are denoted as $(a_j, \mu_j, \sigma_j, j=1,3)$, where $a_j$, $\mu_j$, and $\sigma_j$ represent the amplitude, mean, and scatter of each Gaussian component $j$, respectively.
The true age of the $i$-th host (${\rm Age}^*_i$) is drawn from the sum of three Gaussians, consisting of the nine parameters mentioned above.
Our model for the regression follows the assumptions below:
\begin{equation}
\begin{array}{c}
   \alpha \sim \text{Uniform}(-1, 1) \\
   \beta \sim \text{Uniform}(-1, 0) \\
   \sigma_{\rm int} \sim \text{HalfNormal}(2) \\
   {\rm Age}^*_i \sim {\rm GMM}(a_j, \mu_j, \sigma_j, j=1,3)\\
   \text{HR}^*_i \sim \text{Normal}(\alpha + \beta \text{Age}^*_i , \sigma_{\rm int}^2) \\
   \text{HR}_i \sim \text{Normal}(\text{HR}^*_i , \sigma_{\text{HR},i}^2),
\end{array}
\end{equation}
where the tilde symbol denotes that the parameters are drawn from the specified distributions on the right side.
We assume that $\alpha$ and $\beta$ are assigned a uniform prior distribution\footnote{We have tested different priors for $\alpha$ and $\beta$ using a $\text{Uniform}(-\infty, \infty)$ distribution, but the effect of these changes on the regressions is almost negligible.} within a specific range presented above and the intrinsic scatter $\sigma_{\rm int}$ is modeled as a latent variable following a positive half-normal distribution.
Based on the above models, we estimate the slope ($\beta$), intercept ($\alpha$), and intrinsic scatter ($\sigma_{\rm int}$) using the PyMC3 NUTS sampler \citep{2016ascl.soft10016S}, with a posterior sample size of 100,000 after discarding the initial 24,000 samples, leaving 76,000 samples for analysis.

Figure~\ref{f7} shows the resulting posterior distributions for the \citetalias{2019ApJ...874...32R} host sample.
The slope after the full consideration of the posterior distribution for host age is $-0.038 \pm 0.007$ ($5.52\sigma$).
Compared to the regression based on the LINMIX, the slope is shallower but the significance of the slope increases from $4.4\sigma$ to $5.5\sigma$.
As presented in Figure~\ref{f8}, the slope for the \citetalias{2011ApJ...740...92G} hosts becomes much shallower, down to $-0.025 \pm 0.006$, but the significance of the regression similarly increases from $3.2\sigma$ to $4.2\sigma$, compared to the regression based on the LINMIX.
This suggests that, despite the presence of some dilution bias in this approach, the robust correlation between host age and HR remains evident \citepalias[see also][]{2021MNRAS.503L..33Z}.

\begin{figure}
\centering
\includegraphics[angle=0,scale=0.405]{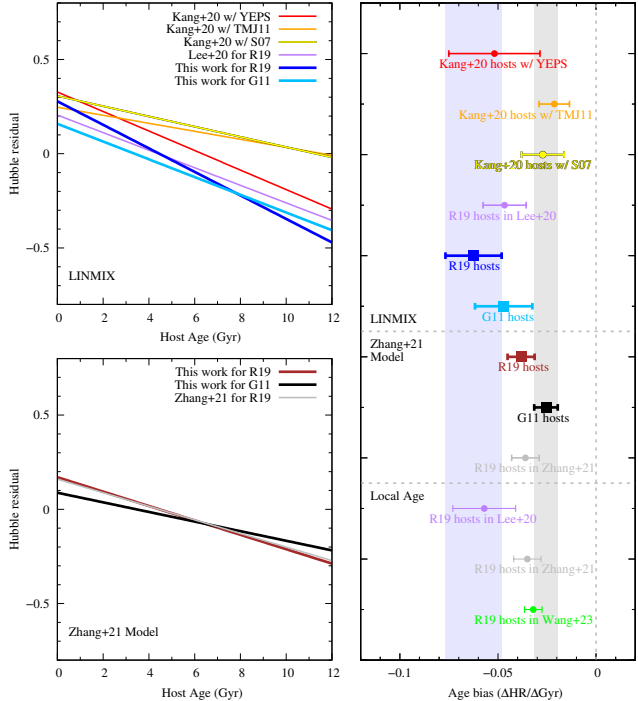}
\caption{
The age bias estimated in various studies to date.
Left column: the top panel compares the linear regressions based on the LINMIX package, while the bottom panel shows the results based on \citetalias{2021MNRAS.503L..33Z} model, using the full posterior distribution for host age derived from MCMC sampling.
All studies show a negative correlation between host age and HR.
Right column: the 68\% credible interval constraints on the age biases reported in various studies along with our new analyses.
The color codes for different studies follow the left column, and the filled squares with thick line indicate this work based on the new age measurement.
The results in the literature based on the local ages are also compared in the bottom section.
The vertical gray dashed line indicates the case of no correlation between host age and HR.
The maximum and minimum credible intervals from our new age measurement are marked with blue and gray shades in the plot.
}
\label{f10}
\end{figure}

Figure~\ref{f9} compares the results of our regression model with those from the LINMIX analysis.
We present a density plot illustrating the probability distribution of HR and the posterior distribution for host age.
Although the regressions using \citetalias{2021MNRAS.503L..33Z} method (orange lines) align more closely with the density maps than LINMIX (dashed gray lines), they appear to slightly deviate in some overdense regions. 
As discussed, this arises from host age posterior distributions with multiple Gaussian components, causing the model to handle complexity differently from empirical density estimation.
Nevertheless, the host age and HR exhibit a negative correlation, indicating that younger hosts tend to be fainter while older hosts are relatively brighter.
As described above, \citetalias{2021MNRAS.503L..33Z} regression models produce shallower slopes compared to those derived from LINMIX, but the significance is considerably increased.
We further present a significance test for a non-positive slope, showing that the probability of a positive trend is nearly zero.
This strongly reinforces the presence of a significant age bias among SN Ia host galaxies.
Even after accounting for the full posterior distribution of host age, age bias remains dominant in our analysis of $\sim 300$ SN Ia hosts.

\section{Comparison with previous analyses} \label{s7}

Figure~\ref{f10} summarizes the SN Ia host age bias reported thus far (\citealt{2020ApJ...889....8K}; \citealt{2020ApJ...903...22L}; \citetalias{2021MNRAS.503L..33Z}; \citealt{2023SCPMA..6629511W}) along with our new analyses for the \citetalias{2011ApJ...740...92G} and \citetalias{2019ApJ...874...32R} samples.
We include the results from \citet{2020ApJ...889....8K} for early-type hosts to illustrate the universality of the age bias in SN Ia hosts, regardless of their morphological types.
We divide the results of the age bias into two categories based on the regression methods: those using LINMIX, which assumes Gaussian age error, and \citetalias{2021MNRAS.503L..33Z} regression model utilizing full posterior for the age error.
The overall correlation presented in the left panels demonstrates the strong ubiquity of the host age bias across various studies, regardless of the regression methods used.

The age bias reported by \citet{2020ApJ...889....8K} shows the impact of different population synthesis models employed in the age measurement.
Due to the small sample size, the significance of the correlation is lower compared to other studies.
However, regardless of the choice of population synthesis models, the age bias persists, ranging between $-0.052$ and $-0.021$ ($\Delta {\rm HR}/\Delta {\rm Gyr}$).
Our previous result for \citetalias{2019ApJ...874...32R} host sample \citep{2020ApJ...903...22L}, based solely on the \citetalias{2019ApJ...874...32R} age measurements, also shows a significant age bias within the range reported by \citet{2020ApJ...889....8K}, at approximately $4.2\sigma$.
As explained in Section~\ref{s5}, our new age measurement reveals stronger age biases in the LINMIX regression for \citetalias{2019ApJ...874...32R} and \citetalias{2011ApJ...740...92G} hosts compared to previous studies by \citet{2020ApJ...903...22L} and \citetalias{2011ApJ...740...92G}.
The strong statistical significance of the correlation is further confirmed in the \citetalias{2021MNRAS.503L..33Z} model applied in the regression.
A persistent strong age bias, regardless of the regression methodology employed, suggests that SNe Ia in younger hosts are significantly fainter than those in older hosts after standardization.
The right panel of Figure~\ref{f10} confirms this trend, showing negative slopes between host age and HR across various studies.

The progenitor ages of SNe Ia are thought to be more closely related to the local environment around the site of SN Ia \citep{2015ApJ...802...20R, 2020A&A...644A.176R, 2018ApJ...854...24K}.
Therefore, using local age measurements to derive progenitor age bias, rather than global age, would offer more precise constraints.
However, the lack of significant difference between the slopes derived from local and global ages suggests that global ages can also serve as reliable tracers of progenitor age.
The bottom section of the right panel in Figure~\ref{f10} shows that the slopes based on the local ages of \citetalias{2019ApJ...874...32R} hosts from different studies remain within the credible bands of their corresponding global age results (\citealt{2020ApJ...903...22L}; \citetalias{2021MNRAS.503L..33Z}).
Note that \citet{2023SCPMA..6629511W} also reported similar slopes in their analysis based on the local age (the green symbol).
With around 300 host ages derived using the same MCMC age sampling, our new analysis provides the most robust evaluation for the progenitor age bias among SN Ia host galaxies.
Including the results presented in this paper, the average slope of all reported age biases and the most conservative value from \citetalias{2011ApJ...740...92G} hosts are $\Delta {\rm HR}/\Delta{\rm Gyr} = -0.033$ and $-0.025$, respectively.

\section{Discussion}\label{s8}

The debate surrounding the luminosity standardization of SN Ia centers on the question: ``What host galaxy property primarily influences variations in SN Ia luminosities after standardization?''.
To address this, we have made reliable host age measurements, spanning from the nearby universe to $z \sim 0.4$, and identified a strong age bias in SN Ia luminosity regardless of host sample selection.
Besides stellar population age, other host properties are also known to be correlated with SN Ia luminosity, such as host mass, metallicity, dust content, and SFR.
In our previous work, we have tested some of these properties, host mass, metallicity, and dust content \citep{2020ApJ...889....8K, 2022MNRAS.517.2697L}.
However, we found only $1.25\sigma$, $0.98\sigma$, and $1.22\sigma$ offsets in the width-luminosity relation (WLR) between the two subgroups based on host mass, metallicity, and dust subgroups, respectively.
This led us to conclude that their impacts on SN Ia luminosity are rather limited compared to the significant impact of stellar population age, which shows a $4.6\sigma$ magnitude offset in the WLR \citep{2022MNRAS.517.2697L}.
\citet{2023ApJ...959...94C} also reported that the root cause of the host mass step, the property most frequently considered in SN Ia standardization, is the stellar population age, based on the empirical galaxy color--magnitude relation.
As well demonstrated in \citet{2022MNRAS.517.2697L}, the origin of the correlation between HR and host age can be understood through the luminosity standardization process, which involves the WLR and CLR that show significant progenitor age dependence \citep[see Figure~2 of][]{2022MNRAS.517.2697L}.
Given this, understanding the origin of the observed step between the local SFR and SN Ia luminosity in nearby hosts is equally important, as SFR is closely related to progenitor age \citep{2015ApJ...802...20R, 2020A&A...644A.176R, 2018ApJ...854...24K}.
We will discuss these issues in detail in Paper III of this series (Park et al. 2025, in preparation), focusing on the relationship between stellar population age and SFR.

The HR is also an important parameter that influences the age bias reported in this study.
Several well-known SN Ia datasets, such as Pantheon+ \citep{2021ApJ...909...26B} and DES-SN5YR \citep{2024arXiv240218690M}, provide consistently estimated HRs for a large sample of SNe Ia.
Using these catalogues to assess the age bias would provide a valuable crosscheck for the universality of the age bias trend.
Unfortunately, unlike Pantheon+, the DES-SN5YR hosts do not overlap with our SN Ia hosts that have measured ages.
In the case of Pantheon+, on the other hand, like many other SN Ia catalogues, employs the host mass step in the standardization process of SN Ia luminosity, even though the root cause of the host mass step is most likely the progenitor ages of host galaxies \citep{2023ApJ...959...94C}.
The typical `down-sizing' phenomenon of observed galaxies \citep[e.g.,][]{2009MNRAS.397.1776F}, characterized by a positive correlation between galaxy age and mass, may include the host age effect within their host mass.
However, our previous studies \citep{2022MNRAS.517.2697L, 2023ApJ...959...94C} indicate that host mass does not fully substitute for progenitor age and reflects only part of the age effect.
Moreover, the Pantheon+ catalogue employed different $R_V$ values of 1.5 and 2.75 based on the host mass threshold of $10^{10} M_\odot$, corresponding to approximately 0.1~mag variation, the typical value for the mass step correction in other SN Ia catalogues. However, this ad-hoc assumption is not supported by observations of $R_V$ values in both high-$z$ galaxies and local galaxies at a given galaxy mass \citep{2018ApJ...859...11S}.
Most critically, applying the mass step correction, instead of the age bias correction, in the SN Ia luminosity standardization regardless of redshift could result in seriously biased outcomes. 
This is because host mass and age evolve differently with redshift. Unlike the insignificant mass evolution of galaxies within $0.0 < z <1.0$, \citep[e.g.,][]{2021MNRAS.503.4413M,2021MNRAS.506.4933A}, the cosmological age evolution during this period accounts for almost half the age of the universe \citep[e.g.,][]{2016JCAP...05..014M, 2022MNRAS.517.2697L, 2022ApJ...927..164B}.

This significant difference in mean progenitor age between the local and high-redshift SNe Ia should have crucial impacts for cosmology. 
As shown in Figure~6 of \citet{2022MNRAS.517.2697L}, while the mean ages of ``young'' progenitors remain around 0.5~Gyr at all redshifts, those for the ``old'' progenitors show an absolute age shift from $\sim$10~Gyrs to $\sim$3~Gyrs with the redshift range from $z \sim 0.0$ to $1.0$.
Figure~\ref{f4} shows that the HR should vary considerably ($\sim$0.2~mag) between these two ages. 
Even when combining the ``young'' and ``old'' progenitors, an average age difference is predicted to be 5-6~Gyrs in this redshift range, which would still result in a substantial change in HR ($\sim$0.15~mag).
However, this potentially important systematic bias has not been included in most previous studies.  
For instance, the DES-SN program, as presented in \citet{2024ApJ...975...86V}, considered only the redshift evolution of the fraction between the ``young'' and ``old'' progenitors, without including the impact of the significant absolute age evolution of ``old'' progenitors across redshift.

Recently, the standard $\Lambda$CDM model faces serious challenges as the observational results from various cosmological probes have become more precise and extensive.
Multiple cosmological probes reveal significant discrepancies, such as the Hubble tension \citep[e.g.,][]{2021CQGra..38o3001D, 2022ApJ...934L...7R} between SN Ia measurements \citep[e.g.,][]{2022ApJ...938..110B} and cosmic microwave background (CMB) results \citep[e.g.,][]{2020A&A...641A...6P}.
More fundamentally, the cosmological principle underpinning the standard cosmological model is also challenged by the discovery of the kinematic dipole of radio quasars \citep[e.g.,][]{2021ApJ...908L..51S, 2023ApJ...953..144G}.
In this context, the SN Ia community's overlooking of the age bias, which would significantly influence the cosmological parameters inferred from SN Ia distance measurements, may have substantial ramifications for the current framework of standard cosmology.
In particular, it is important to note that the recent result from the DESI barionic acoustic oscillation project \citep{2024arXiv240102929D}, coupled with the CMB data \citep{2020A&A...641A...6P}, suggests a time-varying dark energy equation of state.
When the SN dataset is further incorporated into these probes, they show $2.5$ to $3.9 \sigma$ tension with the $\Lambda$CDM model, depending on the SN dataset used.
They predict that the future universe will transition to decelerated expansion although the current universe is in a state of accelerated expansion.
However, these results were derived without considering the progenitor age bias effect in SN cosmology.
Our Paper II of this series (Son et al. 2025, in preparation) will discuss this issue, fully taking into account the impact of the age bias.

\section*{Acknowledgments}
We thank the referee for helpful suggestions.
We acknowledge support from the National Research Foundation of Korea to the Center for Galaxy Evolution Research (RS-2022-NR070872, RS-2022-NR070525).

\section*{Data Availability}

The complete table of new age measurements for \citetalias{2011ApJ...740...92G} and \citetalias{2019ApJ...874...32R} is available as online supplementary material.












\bsp	
\label{lastpage}
\end{document}